\documentclass{pasj01}
\draft
\usepackage{lscape}
\begin{document}
\title{Cloud-cloud Collision in the Galactic Center\\ 50 km s$^{-1}$ Molecular Cloud}
\Received{$\langle$12-Dec-2014$\rangle$}
\Accepted{$\langle$29-Jul-2015$\rangle$}
\Published{$\langle$publication date$\rangle$}
\author{Masato Tsuboi$^{1,2}$,  Atsushi Miyazaki$^3$, and Kenta Uehara$^2$}%
\altaffiltext{1}{Institute of Space and Astronautical Science, Japan Aerospace Exploration Agency,\\
3-1-1 Yoshinodai, Chuo-ku, Sagamihara, Kanagawa 252-5210, Japan }
\email{tsuboi@vsop.isas.jaxa.jp}
\altaffiltext{2}{Department of Astronomy, the University of Tokyo, Bunkyo, Tokyo 113-0033, Japan}
\altaffiltext{3}{Department of Applied Informatics, Hosei University, Kajino, Koganei, Tokyo 184-8584, Japan}
\maketitle

\begin{abstract}
We performed a search of star-forming sites influenced by external factors, such as  SNRs, HII regions, and cloud-cloud collisions,  to understand the star-forming activity in the Galactic center region using the NRO Galactic Center Survey in SiO $v=0, J=2-1$, H$^{13}$CO$^+ J=1-0$, and CS $J=1-0$ emission lines obtained by the Nobeyama 45-m telescope.
We found a half-shell like feature (HSF) with  a high integrated line intensity ratio of $\int T_{\mathrm B}$(SiO $v=0, J=2-1$)$dv$/$\int T_{\mathrm B}$(H$^{13}$CO$^+ J=1-0$)$dv\sim6-8$ in the 50 km s$^{-1}$ molecular cloud, which is a most conspicuous molecular cloud in the region and harbors an active star-forming site seen as several compact HII regions. The high ratio in the HSF indicates that the cloud contains huge shocked molecular gas. The HSF is also seen as a half-shell feature in the position-velocity diagram. A hypothesis explaining the chemical and kinetic properties of the HSF is that the feature is originated by a cloud-cloud collision (CCC).  
We analyzed the CS $J=1-0$ emission line data obtained by Nobeyama Millimeter Array to reveal the relation between the HSF and the molecular cloud cores in the cloud. We made a cumulative core mass function (CMF) of the molecular cloud cores within the HSF. The CMF in the CCC region is not truncated at least up to $\sim2500M_\odot$ although the CMF of the non-CCC region reaches the upper limit of $\sim1500M_\odot$. Most massive molecular cores with  $M_{\mathrm{gas}}>750 M_{\odot}$ are  located only around the ridge of the HSF and adjoin the compact HII region. These may be a sign of massive star formation induced by CCC in the  Galactic center region. 
\end{abstract}

\section{Introduction}
The Central Molecular Zone (CMZ) (\cite{MorrisSerabyn}) is  the Galactic counterpart of the central molecular cloud condensation often observed in nearby spiral galaxies (e.g. \cite{Kazushi}). The molecular clouds in the CMZ are much denser, warmer, and more turbulent than those in the disk. There are young and highly luminous star clusters in the CMZ, for example Arches cluster and  Quintuplet cluster (e.g. \cite{Figer1999}, \cite{Figer2002}). The star formation in the CMZ must be influenced by external factors, such as interactions with SNRs, HII regions, and cloud-cloud collisions because they are crowded in the region (e.g. \cite{Morris1993}, \cite{Hasegawa1994},  \cite{Hasegawa2008}). However, it is hard to know observationally how the cradle molecular clouds produce such massive clusters because these clusters have already lost the surrounding molecular materials. 

There are two most conspicuous molecular cloud surrounding the Galactic Center itself,  Sagittarius A$^{\ast}$(Sgr A$^{\ast}$). They are called the 20  and 50 km s$^{-1}$ molecular clouds after their LSR radial velocities (hereafter, 50MC and 20MC).   The 50MC is located only $3^{\prime}$ from Sgr A$^{\ast}$. The 50MC probably harbors a massive active star-forming site which is seen as several compact HII regions (e.g. \cite{Ekers1983}, \cite{Goss1985}, \cite{Yusef-Zadeh2010}, \cite{Mills}) although the 20MC is not associated with obvious HII regions.  The star formation activity of the 50MC is on-going and can be spatially resolved by radio and IR observations. In addition, the 50MC is probably associated with a young SNR, Sgr A East (e.g. \cite{Tsuboi et al. 2009}). Therefore the 50MC is the best laboratory for understanding star formation influenced by external factors in the CMZ. Moreover it may be the nearest analog of star formation in extragalactic nuclei.

In the third section, we search the evidence of interactions with SNRs, HII regions, and cloud-cloud collisions in the Sgr A region, which is a main part of the CMZ, using the Galactic Center  Survey  
with the 45-m telescope at the Nobeyama Radio Observatory (NRO)\footnote{The Nobeyama Radio Observatory is a branch of the National Astronomical Observatory, National Institutes of Natural Sciences, Japan.}. We found a peculiar feature in the 50MC.
In the fourth section, we make core mass function (CMF) in the cloud using the CS $J=1-0$ emission line data with Nobeyama Millimeter Array (NMA) and explore the relation between the peculiar feature and the molecular cloud cores in the cloud. Throughout this paper, we adopt 8.5 kpc as the distance to the Galactic center. Then, $24\arcsec$ corresponds to about 1 pc at the distance. And we use Galactic coordinates.
\begin{figure}
\begin{center}
\includegraphics[width=11cm]{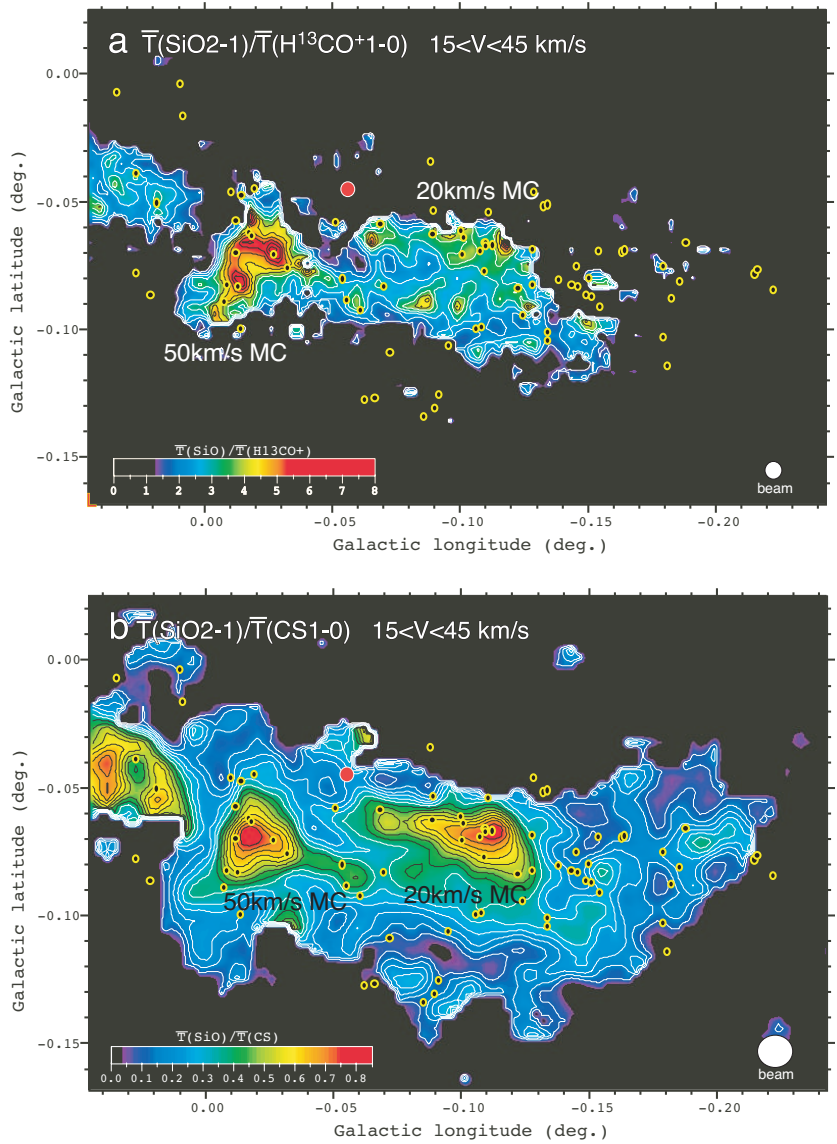}
\end{center}
\caption{{\bf a} The 20 and 50 km s$^{-1}$ molecular clouds in an integrated line intensity ratio of 
$\int T_{\mathrm B}$(SiO $v=0, J=2-1$)$dv$/$\int T_{\mathrm B}$(H$^{13}$CO$^+ J=1-0$)$dv = \bar{T}_{\mathrm B}$(SiO$2-1$)/$\bar{T}_{\mathrm B}$(H$^{13}$CO$^+1-0$). The integrated velocity range is $V_{\mathrm LSR}=15-45$ km s$^{-1}$. The angular resolution is $26{\arcsec}$. Red circle shows the position of Sgr A$^{\ast}$. Yellow circles show the positions of CH$_3$OH Class I maser at 36 GHz (\cite{Yusef2013}). {\bf b}  The 20 and 50 km s$^{-1}$ molecular clouds in an integrated line intensity  line ratio of $\bar{T}_{\mathrm B}$(SiO$2-1$)/$\bar{T}_{\mathrm B}$(CS$1-0$).  The  velocity range is $V_{\mathrm LSR}=15-45$ km s$^{-1}$. The angular resolution is $60{\arcsec}$.  ~~}
\label{Fig1}
\end{figure}
\section{Observations}
We used our observational datasets (channel maps) with the NRO 45-m telescope and NMA, which have been already published as other papers. A short summary of our observations is as follows; 
We have observed the Sgr A region in CS $J=1-0$ (48.990964 GHz)  by using the NRO 45-m telescope (\cite{Tsuboi1999}). We also have observed the region in SiO $J=2-1$ ($\nu=86.847010$GHz) and H$^{13}$CO$^+$ $J=1-0$ ($\nu=86.754330$GHz) emission lines by using the NRO 45-m telescope (\cite{Tsuboi2011}). These are parts of the  NRO Galactic Center  Survey. The velocity resolution of these channel maps was 5.0 km s$^{-1}$. The angular resolutions were $60\arcsec$ for the CS $J=1-0$ maps and $26\arcsec$ for the SiO $J=2-1$ and H$^{13}$CO$^+ J=1-0$ maps, respectively. These correspond to about $2.5$ pc and $1.1$ pc at the distance to the Galactic center, respectively.

We have observed the 50MC in CS $J=1-0$ emission line by using NMA (\cite{Tsuboi et al. 2009}). The FWHM of the element antenna of NMA is $156\arcsec$ at 49 GHz. The velocity resolution of the channel map was 3.8 km s$^{-1}$. To make maps, we used the CLEAN method in the NRAO AIPS package. The size of the synthesized beam was $8.5\arcsec\times10\arcsec (\phi=24^\circ)$ for a natural weighting, which corresponds to about $0.35$ pc $\times 0.42$ pc at the distance to the Galactic center. Features with spatial scales larger than $1'$ or 2.5 pc were resolved out.

\section{Cloud-Cloud Collision in the 50 km s$^{-1}$ Molecular Cloud}
\subsection{Huge Shocked Molecular Gas in the  50 km s$^{-1}$ Molecular Cloud}
The SiO $J=2-1$ emission line is strongly detected toward the whole 50MC (see Fig.3 in \cite{Tsuboi2011}).  Si atom is sputtered from dust grains by non-dissociatve C-shocks, which later in the gas-phase forms SiO molecule (e.g. \cite{Ziurys et al 1989}, \cite{Usero}, \cite{Amo}). The velocity width required for the ejection is $\Delta v\gtrsim 30$ km s$^{-1}$. The SiO emission line is therefore commonly accepted to be a tracer of shocked molecular gas. The area with enhancement of SiO molecule indicates the region where shock wave propagated within $10^5$ yr because the molecule is adsorbed again on dust grain during the time scale. Meanwhile the fractional abundance of H$^{13}$CO$^+$ ion is not enhanced by C-shock. The SiO $J=2-1$ and H$^{13}$CO$^+$ $J=1-0$ ($\nu=86.754330$GHz) emission lines are observed simultaneously because the frequency separation between the emission lines is only 93 MHz. They are optically thin even in the Galactic center molecular cloud (e.g. \cite{Handa2006}).  
Therefore the integrated line intensity ratio, $R_{\mathrm{SiO/H}^{13}\mathrm{CO}^+} = \int T_{\mathrm B}$(SiO $v=0, J=2-1$)$dv$/$\int T_{\mathrm B}$(H$^{13}$CO$^+ J=1-0$)$dv = \bar{T}_{\mathrm B}$(SiO$2-1$)/$\bar{T}_{\mathrm B}$(H$^{13}$CO$^+1-0$), is suitable to search for shocked molecular gas avoiding the effect of  the difference of the observation conditions.

Figure 1a shows the Sgr A region in an integrated line intensity ratio of $R_{\mathrm{SiO/H}^{13}\mathrm{CO}^+}$.  
The integrated velocity range of the figure is $V_{\mathrm LSR}=15-45$ km s$^{-1}$. The angular resolution is $26{\arcsec}$. We identify a half-shell like feature with $R_{\mathrm{SiO/H}^{13}\mathrm{CO}^+}\sim6-8$ in the 50MC (hereafter, HSF). The diameter of the HSF is $D\sim 3$ pc. Meanwhile, there is not such feature in the 20MC.
The enhancement of the ratio must be caused by the increase of the fractional abundance of SiO molecule and/or the decrease of the fractional abundance of H$^{13}$CO$^+$ molecule. 
The 50MC is probably associated with a massive active star-forming site as mentioned in the Introduction.  
H$^{13}$CO$^+$ ion may be destroyed by strong UV radiation from the young massive stars. 
The decrease of the fractional abundance of H$^{13}$CO$^+$ ion  around the star-forming site  produces an apparent circle feature with a high ratio of $R_{\mathrm{SiO/H}^{13}\mathrm{CO}^+}$ even when  the fractional abundance of SiO molecule does not increase. In this case, there is still an open question which is a dominant factor because  any obvious sign of the feature is not seen both in the channel maps of SiO $v=0, J=2-1$ and H$^{13}$CO$^+ J=1-0$ emission line (see Fig.4 in \cite{Tsuboi2011}).  

On the other hand, CS molecule is not so significantly enhanced by C-shock and is not so easy destroyed by UV radiation (see Fig.6 in \cite{Mart}). An integrated line intensity ratio of $R_{\mathrm{SiO/CS}}= \int T_{\mathrm B}$(SiO $v=0, J=2-1$)$dv$/$\int T_{\mathrm B}$(CS $J=1-0$)$dv$ is also an indicator of shocked molecular gas, which probably has  insensibility to UV radiation.  The angular resolution of the data is $60{\arcsec}$.  Figure 1b shows the Sgr A region in the integrated line intensity ratio. The integrated velocity range is the same as that of figure 1a.
We identify a prominent component with the high ratio of $R_{\mathrm{SiO/CS}}$ in the 50MC.  The integrated line intensity ratio is $R_{\mathrm{SiO/CS}}=0.8\pm0.05$ at the center position, which is three times larger than the ratio, $R_{\mathrm{SiO/CS}}\sim0.3$, in other part of the 50MC.  Although the HSF clearly seen in figure 1a is almost concealed by the lower angular resolution of the CS data ($60{\arcsec}$), it is barely identified as a curved  feature in figure 1b.  There is a possibility that the optical thickness of the CS emission line is somewhat large in the center region of the molecular cloud and produce a feature with a high ratio without the increase of the fractional abundance of SiO molecule. However, the half-shell morphology is hard to be  produced by this effect because the optical thickness at the center position should be larger than that of the surrounding region. In addition, the integrated line intensity ratio of $R_{\mathrm{H}^{13}\mathrm{CO}^+/\mathrm{CS}} = \int T_{\mathrm B}$(H$^{13}$CO$^+$)$dv$/$\int T_{\mathrm B}$(CS)$dv$ is uniformly $\sim0.10-0.15$ in the whole 50MC.  There is no feature with a peculiar integrated line intensity ratio in the cloud.   
These rule out the possibility that the optical thickness of the CS emission line produces the feature apparently. Then these indicate that the  50MC contains huge shocked molecular gas. 

There is another feature with the high ratio in the north part of the 20MC in figure 1b. The integrated line intensity ratio is $R_{\mathrm{SiO/CS}}=0.8\pm0.05$ at the peak position, which is larger than the ratio, $R_{\mathrm{SiO/CS}}\simeq0.3\pm0.1$, in the south part of the 20MC.  Meanwhile this feature has no significant counter part with a high ratio of $R_{\mathrm{SiO/H}^{13}\mathrm{CO}^+}$  in figure 1a.
Yellow circles in figures 1a and 1b show the positions of Class I methanol (CH$_3$OH) maser at  36.17 GHz (\cite{Yusef2013}). Class I methanol masers are understood to be collisionally pumped in shocks. The positions of the maser are associated with the feature. This suggests that the  production of the shocked molecular gas is on-going in the feature. The feature probably contains large amount of shocked molecular gas. In addition, there is another group of the masers around $l\sim -0.15^\circ~b\sim-0.08^\circ$. This may be associated with another SiO feature with lower LSR velocity (Fig.8 in \cite{Tsuboi2012}). 

The difference of $R_{\mathrm{SiO/H}^{13}\mathrm{CO}^+}$ between the features associating with the 20MC and 50MC may be caused by the strength of the shock.
As mentioned previously, the production of the SiO molecule requires the velocity width of $\Delta V\gtrsim30$  km s$^{-1}$ and it becomes more active with increasing of the velocity width. Typical velocity widths of the SiO and H$^{13}\mathrm{CO}^+$ emission lines in the 20MC are narrower than those in the 50MC (see Fig.2 in \cite{Tsuboi2011}).

\begin{figure}
\begin{center}
\includegraphics[width=11cm]{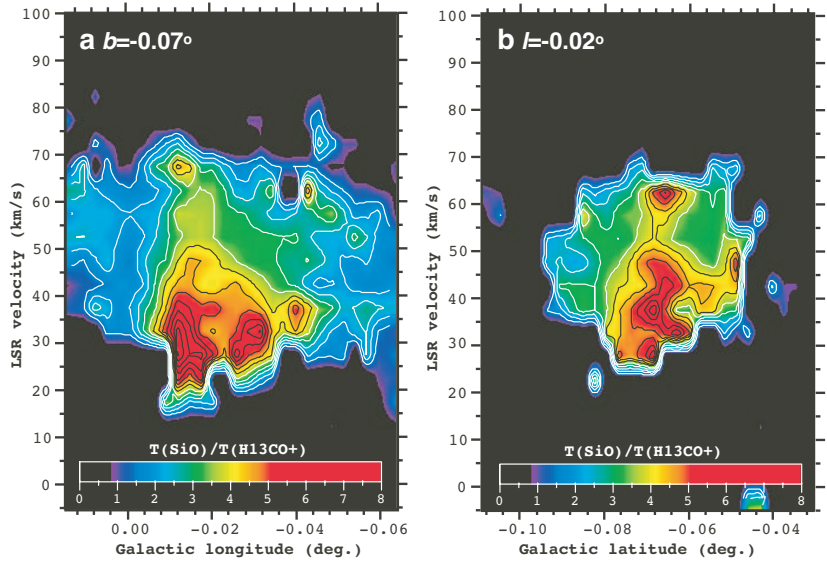}
\end{center}
\caption{Position-velocity daigrams of shocked molecular gas  in the  50 km s$^{-1}$ molecular cloud traced by $T_{\mathrm B}$(SiO)/$T_{\mathrm B}$(H$^{13}$CO$^+$). {\bf a} Galactic longitude- velocity diagram along $b=-0.07^{\circ}$. {\bf b} Galactic latitude- velocity diagram along $l=-0.02^{\circ}$.}
\label{Fig2}
\end{figure}

\subsection{Origin of Shocked Molecular Gas in the  50 km s$^{-1}$ Molecular Cloud}
Figure 2 shows the kinematic structures of the HSF traced by an  integrated line intensity ratio of  $R_{\mathrm{SiO/H}^{13}\mathrm{CO}^+}$. Figure 2a is the Galactic longitude- velocity diagram along $b=-0.07^{\circ}$. The HSF seen in figure 1a is also identified as a half-shell like feature  with a high ratio of $ R_{\mathrm{SiO/H}^{13}\mathrm{CO}^+}\sim5-8$  in the diagram. 
The velocity expanse of the feature, full width at zero-intensity (FWZI), is as large as $\Delta V_\mathrm{FWZI}\sim50$ km s$^{-1}$. 
This is consistent with the velocity width required for the production of SiO molecule. Figure 2b is the Galactic latitude- velocity diagram along $l=-0.02^{\circ}$.  A feature with a high ratio corresponding the HSF is also seen in the Galactic latitude- velocity diagram. 

As mentioned previously, no extended HII region is detected in the 50MC although several compact HII regions are detected. At least, young massive stars in the disk region have no thermal extended SiO  emission comparable to the HSF. Thermal SiO emission line are detected around a massive young star in the disk region, the Orion source I (\cite{Niederhofer}). The spatial expanse of the emission line is restricted around the source I, which is less than 0.1 pc or much smaller than the distance between the compact HII regions and the HSF. The velocity expanse of the emission line is as large as 20 km s$^{-1}$, which is smaller than that of the HSF.   In addition, although many massive young stellar objects are detected in the Sgr A region (\cite {An}), they have no clear association with enhanced areas of thermal SiO emission line (\cite{Tsuboi2011}). It is hard to explain the origin of the shocked molecular gas in the 50MC by neighboring HII region. 

Figure 3a shows the positional relation between the HSF in the 50MC and the Sgr A east shell in 5-GHz radio continuum observed by VLA (contours; \cite{Yusef-Zadeh1987}).  The northwestern boundary of the HSF coincides with the southeastern boundary of the Sgr A east shell.  An extended component with moderate integrated line intensity ratio, $R_{\mathrm{SiO/H}^{13}\mathrm{CO}^+} \sim3-4$, traces the southeastern boundary of the SNR shell. The integrated line intensity ratio probably shows that the boundary region contains shocked molecular gas. However, most of the HSF is located outside of the southeastern boundary of the SNR shell although this feature contains huge shocked molecular gas as mentioned above. The positional anti-correlation suggests that the HSF was not produced by the interaction with the SNR shell.

These kinematic features shown in Figure 2 and the appearance shown in Figure 1 indicate that the shocked molecular gas has a hollow hemisphere-like shape in the $l-b-v$ space. The one-sided feature in the $l-b-v$ space suggests that  something  impacts on one side of the molecular cloud and the induced C-shock produces huge shocked molecular gas in the cloud. The shape in the $l-b-v$ space of the shocked molecular gas are consistent with simulations of cloud-cloud collision (CCC) (e.g.\cite {Habe}, \cite{Takahira}). In the case, 
The velocity expanse of the feature may be as large as the velocity difference between an impactor and a cloud.
If so, this is enough to produce SiO molecule by shock. The dynamical time scale is  $t= D/\Delta V_\mathrm{FWZI}\sim 3 \mathrm{pc}/50\mathrm{km s^{-1} }=6\times10^4$ yr. This is shorter than the life time of SiO molecule in gas phase, $10^5$ yr.  Therefore, a probable hypothesis explaining the observed chemical and kinematic features is that the HSF is produced by CCC.

\begin{figure}
\begin{center}
\includegraphics[width=16cm]{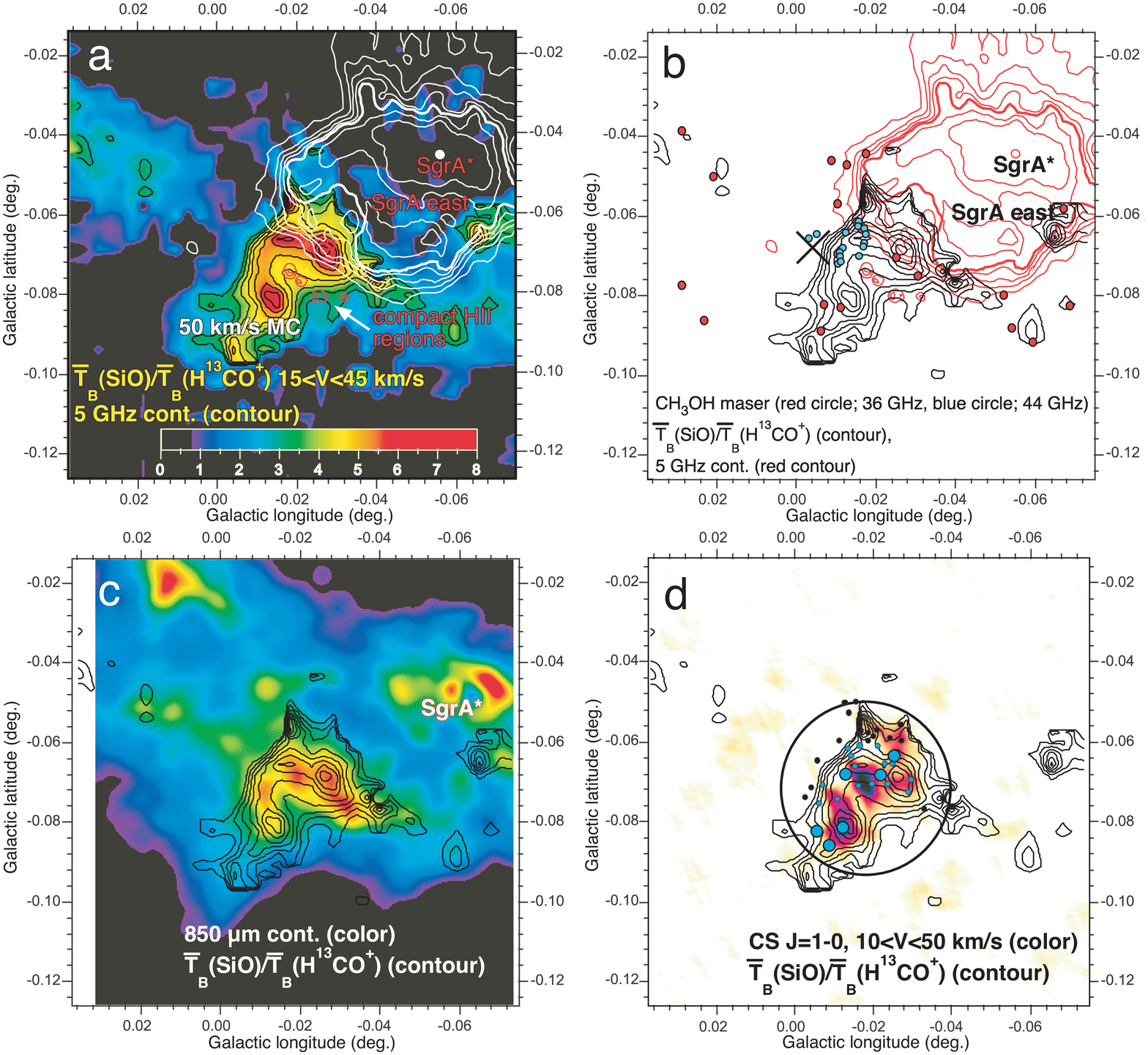}
\end{center}
\caption{{\bf a} Relation between the  50 km s$^{-1}$ molecular cloud (50MC) in $\bar{T}_{\mathrm B}$(SiO$2-1$)/$\bar{T}_{\mathrm B}$(H$^{13}$CO$^+1-0$)  (pseudo color) and the Sgr A east (SgrAE; white contours) in 5-GHz radio continuum with VLA (\cite{Yusef-Zadeh1987}).  The compact HII regions are also shown (red contours; \cite{Yusef-Zadeh2010}). The integrated velocity range of the integrated line intensity ratio is $V_{\mathrm LSR}=15-45$ km s$^{-1}$. {\bf b} Blue circles shows the positions of 44 GHz class I CH$_3$OH masers (\cite{Phist}).   
Red circles shows the positions of 36 GHz class I CH$_3$OH masers (\cite{Yusef2013}) for comparison. 
Cross indicates the reference position where very high gas temperature $>300$ K, has been reported based on NH$_3$ observations (\cite{MillsMorris}). 
{\bf c} Relation between the 50MC in $\bar{T}_{\mathrm B}$(SiO$2-1$)/$\bar{T}_{\mathrm B}$(H$^{13}$CO$^+1-0$) (black contours) and in 850 $\mu$m continuum with JCMT (pseudo color; \cite{Pierce-Price2000}). {\bf d} Filled circles shows the positions of  identified molecular cloud cores overlaid on the integrated intensity map of the CS $J=1-0$ emission line with NMA (\cite{Tsuboi et al. 2009}). Molecular cloud cores which are probably included in the half-shell like feature are shown as blue circles. Large blue symbols show massive molecular cores with  $M_{\mathrm{gas}}>400 M_{\odot}$.  Other molecular cloud cores are shown as black circles. A large circle shows the full width at half-maximum (FWHM) of the element antenna of the NMA at 49 GHz. }
\label{Fig3}
\end{figure}
Figure 3b shows the positions of Class I methanol (CH$_3$OH) maser spots at 36.17 GHz (red circles; \cite{Yusef2013}) and 44.07 GHz (blue circles; \cite{Phist})  observed by  VLA in the the 50MC.   The methanol maser spots at 36 GHz are probably located both around the HSF and the southeastern boundary to the Sgr A east shell.
Meanwhile the methanol maser spots at 44 GHz are probably located  on the HSF itself rather than the contact surface  to the Sgr A east shell.  
The methanol masers at 44 GHz appear to concentrate around the northeastern boundary of the HSF. However, since the SE boundary  out of the full width at half-maximum (FWHM) of the element antenna of VLA at 44 GHz, we cannot rule out a distribution of 44 GHz masers in the SE boundary either.
The Class I methanol maser at 44 GHz is also pumped collisionally by shock wave. However, the methanol maser at 44 GHz requires more extreme conditions for number density and temperature than those of the methanol maser at 36 GHz (\cite{Phist}). They probably indicate the position of propagating shock wave at present.  The concentration of the methanol maser spots at 44 GHz around the northeastern boundary of the HSF supports that the HSF was originated  by  CCC. 
In addition, cross in the figure indicates the center position of the area with very high gas temperature, $T>300$ K, which is detected in the observation of NH$_3$ methastable lines (\cite{MillsMorris}).  This area is probably associated with the concentration of the methanol maser spots at 44 GHz.

Figure 3c shows the positional relation between the 50MC in $R_{\mathrm{SiO/H}^{13}\mathrm{CO}^+}$ and in 850 $\mu$m continuum emission with JCMT (\cite{Pierce-Price2000}), which indicates the distribution of 
dust. The 850 $\mu$m continuum emission traces the HSF very well. The sub-mm flux is proportional to the product of the dust temperature and column density of the dust assuming that the emission is operatically thin. We also assume that the gas-to-dust ratio is not change in the cloud. If the morphology of 850 $\mu$m continuum is mainly controlled by the column density change  of the dust rather than the dust temperature change, the sub-mm morphology should resemble that of the molecular cloud itself, which are traced by CS or ${\mathrm{H}^{13}\mathrm{CO}^+}$ emission lines, rather than that of  $R_{\mathrm{SiO/H}^{13}\mathrm{CO}^+}$.  On the other hand, if the morphology were to be controlled by the dust temperature change, then the sub-mm morphology would trace the site of heating of the dust. This suggests the shock heating of the dust by a CCC.

 \section{Molecular Cloud Cores in the Cloud-cloud Collision Spot}
 \subsection{Identfication of Molecular Cloud Cores}
We used the {\it clumpfind} algorithm (\cite{Williams1994}) to automatically identify the molecular cores in the 50MC. The channel maps, the detail procedure of core identification, and the list of identified molecular cloud core already have been shown in the previous papers (\cite{Tsuboi et al. 2009}, \cite{Tsuboi2012b}).  
In the case of  molecular cores in the disk region, the finding method based on optically thin dust emission observation is seem to be superior to that based on emission line observation (e.g. \cite{Pineda}). However, the Galactic center molecular clouds are too crowded to suffer from confusion when using the former method. The later method  is barely available because this separates the cores with the different radial velocity to the different channel maps.  

We analyzed the molecular cloud core list to explore any influence of the CCC on star formation in the 50MC. Figure 3d shows the position of the molecular cloud cores  overlaid on the integrated intensity map of the CS $J=1-0$ emission line with NMA (\cite{Tsuboi et al. 2009}). Molecular cloud cores which are located in the HSF (black contour) are shown as blue circles. We identified 21 cores in the HSF. We supposed that these molecular cloud cores are contained physically in the spot of the CCC. Molecular cloud cores which are located outside of the HSF are shown as black circles in the figure. They mostly belong to  the interacting region with the Sgr A east shell. 
The positions of the molecular cloud cores and the judgements of whether they are located in the spot of the CCC or not are summarized in Table 1. 
We use "$R_{\mathrm{SiO/H}^{13}\mathrm{CO}^+} > 4$ or $< 4$ at the center position of the core" as the criterion of the judgement because  it is hard to find $R_{\mathrm{SiO/H}^{13}\mathrm{CO}^+} > 4$ in the area except for the HSF  (see Fig.1 and Fig.3a). 
The radial velocities, peak brightness temperatures with the correction of the primary beam attenuation of the NMA,  half width at half-maximum (HWHM)  radius and FWHM velocity widths (also see \cite{Tsuboi2012b}) are also summarized in Table 1.  
To estimate the HWHM radius, we subtracted the physical resolution from the observed radius, $R_o$: 
$R=(R_{\rm o}^2-r_{\rm{beam}}^2)^{1/2}$, where $r_{\rm{beam}}$ is the physical resolution; $r_{\rm{beam}}=\frac{\sqrt{0.35\times0.45}}{2}=0.2$ pc.

 \subsection{Mass Estimation of Molecular Cloud Cores}
The molecular gas mass of the identified cores is re-estimated from the CS $J=1-0$ emission line intensity under the LTE condition. The molecular column density traced by the CS $J=1-0$ emission line with the optical thickness of $\tau$ is given by
\begin{equation}
\label{ }
N_{\mathrm{mol}}(\tau)[\mathrm{cm}^{-2}]=\frac{7.55 \times 10^{11}T_{\mathrm{ex}}\int T_{\mathrm{B}}dv[\mathrm{K~km s}^{-1}]} {X({\mathrm{CS}})}  \times\frac{\tau}{1-e^{-\tau}}.
\end{equation}
Here, $\tau$ is the optical thickness of the CS $J=1-0$ emission line; $X({\mathrm{CS}})$ is the fractional abundance, $X({\mathrm{CS}})=N({\mathrm{CS}})/N_{mol}$, which is the relative abundance of CS molecules to total molecules; $T_{\mathrm{ex}}$ is the excitation temperature of the CS $J=1-0$ emission line. These parameters are likely to be different from core to core.
The $\tau$ in the 50MC had been estimated to be $\tau \sim 1-3$ using the comparison between the emission line intensities of the major and minor isotopes (\cite{Liu}). 
Here, we assumed that the molecular cores have a single optical thickness of  $\tau=1.7$. If so, the remaining error by the optical thickness should be suppress to be less than $\pm50$\%.
Using RADEX LVG program (\cite{van der Tak}) based on CS $J=1-0$ and $2-1$ emission line data in the 50MC (CS $J=1-0$; \cite{Tsuboi1999}, CS $J=2-1$; \cite{Tsuboi1997}, \cite{Bally1987}), the $T_{\mathrm{ex}}$ and $X({\mathrm{CS}})$ of the cores are evaluated. We assumed that the molecular cores have the single gas kinetic temperature of $T_{\mathrm{K}}=80$ K (e.g \cite{MillsMorris}). Figure 4 shows the calculated relations between the brightness temperature ratio of the CS $J=1-0$ and CS $J=2-1$ emission lines,  $T_{\mathrm{B}}$(CS $J=2-1$)/$T_{\mathrm{B}}$(CS $J=1-0$) and the brightness temperature of the CS $J=1-0$ emission line   on the plane of H$_2$ number density, $n_{\mathrm H_2}$, versus CS fractional abundance per velocity gradient, $X$(CS)$/(dv/dr)$. The figure also shows the curves of the excitation temperature of CS $J=1-0$ emission line on the plane of  $n_{\mathrm H_2}$ versus $X$(CS)$/(dv/dr)$. Filled circles in the figure show the identified CS molecular cores in the 50MC. They have large scatters of  $T_{\mathrm{ex}}=20-80$ K and $X(\mathrm{CS}) = 0.5-9\times 10^{-8}$, respectively.  These corrections are applied to the LTE masses. They are also summarized in Table 1.
\begin{figure}
\begin{center}
\includegraphics[width=11cm]{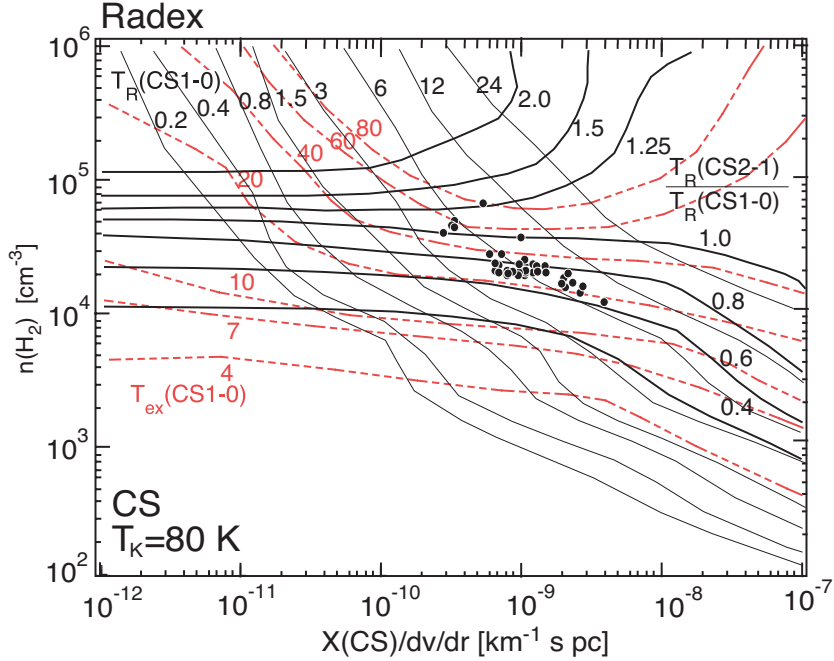}
\end{center}
\caption{The curves of the CS  $J=1-0$ brightness temperatures, $T_{\mathrm{B}}$(CS $J=1-0$) (black thin lines) and the brightness temperature ratio, $T_{\mathrm{B}}$(CS $J=2-1$)/$T_{\mathrm{B}}$(CS $J=1-0$)  (black  thick lines) are shown on the plane of H$_2$ number density vs.~CS fractional abundance per velocity gradient.  The curves are calculated by the RADEX program at a kinetic temperature of $T_\mathrm{K}=80$ K. Filled circles show the identified CS molecular cores in the 50MC. The curves of the excitation temperature  (red broken lines), $T_{\mathrm{ex}}$(CS $J=1-0$) are also shown. }
\label{Fig4}
\end{figure}

We calculate LTE molecular gas mass of the core  from the derived distribution of the molecular column density. The molecular gas mass  is given by
\begin{equation}
\label{eq:2}
M_{\mathrm{gas}}[M_{\odot}] =\Omega[\mathrm{cm}^2]\mu[M_{\odot}] \sum_m\sum_n 
 N_{\mathrm{mol}}(m,n,\tau)[\mathrm{cm}^{-2}],
\end{equation}
where $\Omega$ is the physical area corresponding to the data grid, $\Omega=6.44 \times 10^{34} \mathrm{cm}^2 $ for a $2\arcsec$ grid spacing of the data cube and $\mu$ is  the mass of one particle with mean molecular weight of 2.3; thus, $\mu=1.94\times 10^{-57} M_{\odot}$. The range and mean of the molecular gas mass are  $M_{\mathrm{gas}}= 1.8\times10^2-6.1\times10^3$ M$_{\odot}$ and $\overline{M}_{\mathrm{gas}}=1.1\times10^3$  M$_{\odot}$, respectively.
The molecular gas masses  are summarized in Table 1.  

We also estimate the molecular gas mass by using the virial theorem assuming  no external pressure and no magnetic field.  The virial theorem mass  of a spherical cloud  with uniform density is nominally given by
\begin{equation}
\label{eq:5 }
M_{\mathrm{vir}}[M_{\odot}]=210\times \Delta V_{1/2}[\mathrm{km s}^{-1}]^2 R[\mathrm{pc}]
\end{equation}
where $ \Delta V_{1/2}$ is the FWHM of a Gaussian emission line profile with the correction of the optical thickness.  The range and mean of the observed velocity width of the cores are $\Delta V_{1/2}=2.9-12.9$ km s$^{-1}$ and $\overline{\Delta V}_{1/2}=7.8\pm2.5$ km s$^{-1}$, respectively.  The range and mean of the virial theorem mass in the CS $J=1-0$ emission line are  $M_{\mathrm{vir}}= 2.4\times10^2-1.5\times10^4$  M$_{\odot}$ and $\overline{M}_{\mathrm{vir}}=5.2\times10^3$  M$_{\odot}$, respectively.  The ratio, $\alpha=M_{\mathrm{vir}}/M_{\mathrm{gas}}$, indicates the relation between the virial theorem mass and the LTE molecular gas mass. The mean of the ratio is $\bar{\alpha}=6.9\pm1.7$. They are also summarized in Table~1. This indicates nominally that the cores are not bounded without any external pressure. Although similar situation has been reported for larger-scale molecular clumps in the CMZ (e.g. \cite{MiyazakiTsuboi}), what is the external pressure for bounding of the cores in the cloud is still an open question.  

\begin{figure}
\begin{center}
\includegraphics[width=11cm]{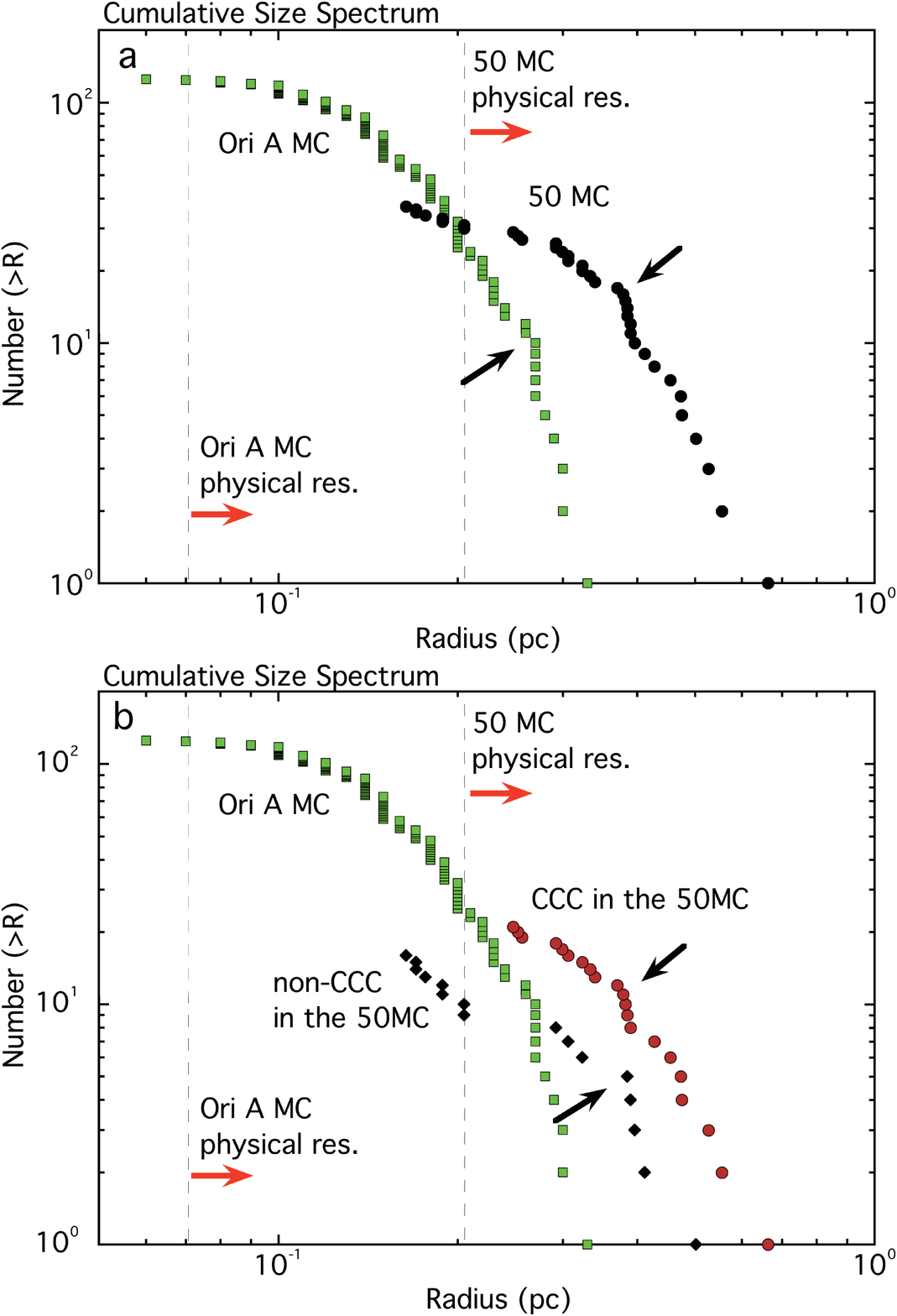}
\end{center}
\caption{{\bf a} Cumulative size spectrum (SS) of the molecular cloud cores within the 50 km s$^{-1}$ molecular cloud (50MC) is shown (black circles). 
The horizontal axis is the HWHM radius  of each clump, $R$. To estimate the HWHM  radius, we subtracted the physical resolution from the observed radius, $R_o$: 
$R=(R_{\rm o}^2-r_{\rm{beam}}^2)^{1/2}$, where $r_{\rm{beam}}$ is the physical resolution; $r_{\rm{beam}}=0.2$ pc. The physical resolution is indicated as a vertical broken line.
The cumulative SS of the Ori A molecular cloud is also shown  for comparison (green squares; \cite{Tatematsu}). The cumulative SSs of the 50MC and OriAMC have breaks at $R\sim0.37$ pc and $R\sim0.25$ pc, respectively (black arrows). 
{\bf b}   Cumulative SSs of the cloud-cloud collision (CCC) (red circles) and non-CCC  (black diamonds) regions in the 50MC are shown. Both SSs have similar breaks at $R\sim0.37$ pc  (black arrows).}
\label{Fig5}
\end{figure}
 \subsection{Size Spectrum of Molecular Cloud Cores}
 Figure 5a shows the cumulative size spectrum (SS) of the molecular cloud cores within the 50 km s$^{-1}$ molecular cloud (50MC). 
The cumulative SS of the Ori A molecular cloud is also shown in the figure for comparison (\cite{Tatematsu}). The 50MC is a factor 20 more distant than the OriAMC, while the effective angular beam size of our data is 6 times sharper than that of the OriAMC data because of the sparse sampling of the OriAMC data. There is still a factor 3 difference between the physical effective resolutions. The physical resolutions are  indicated as vertical broken lines in the figure. 
The cumulative SSs of the 50MC and OriAMC have breaks at $R\sim0.37$ pc and $R\sim0.25$ pc, respectively. The truncations generally means that a radius of the molecular cloud cores reaches any upper limit. The mean radius of the detected cores in the 50MC, $\bar{R}=0.34\pm0.12$ pc, is twice larger than that of the OriAMC data, $\bar{R}=0.16\pm0.06$ pc (\cite{Tatematsu}). Because the values are larger than the physical resolutions, they are not significantly affected by the limitations of the resolution. Therefore the radius of the cores in the 50MC is larger than that in the OriAMC.
  
Figure 5b shows cumulative SSs of the cloud-cloud collision (CCC) and non-CCC regions in the 50MC. Both SSs have similar break at $R\sim0.37$ pc. The mean radiuses of the detected cores in the CCC and non-CCC regions are $\bar{R}=0.39\pm0.11$ pc and $\bar{R}=0.28\pm0.11$ pc, respectively. The radius of the cores in the CCC region is slightly larger than that in the non-CCC region.
However, there is  a possibility that the difference of the radius is produced as artifact by the limitation of the resolution because the half number of cores in the non-CCC region are around the effective resolution.
\begin{figure}
\begin{center}
\includegraphics[width=11cm]{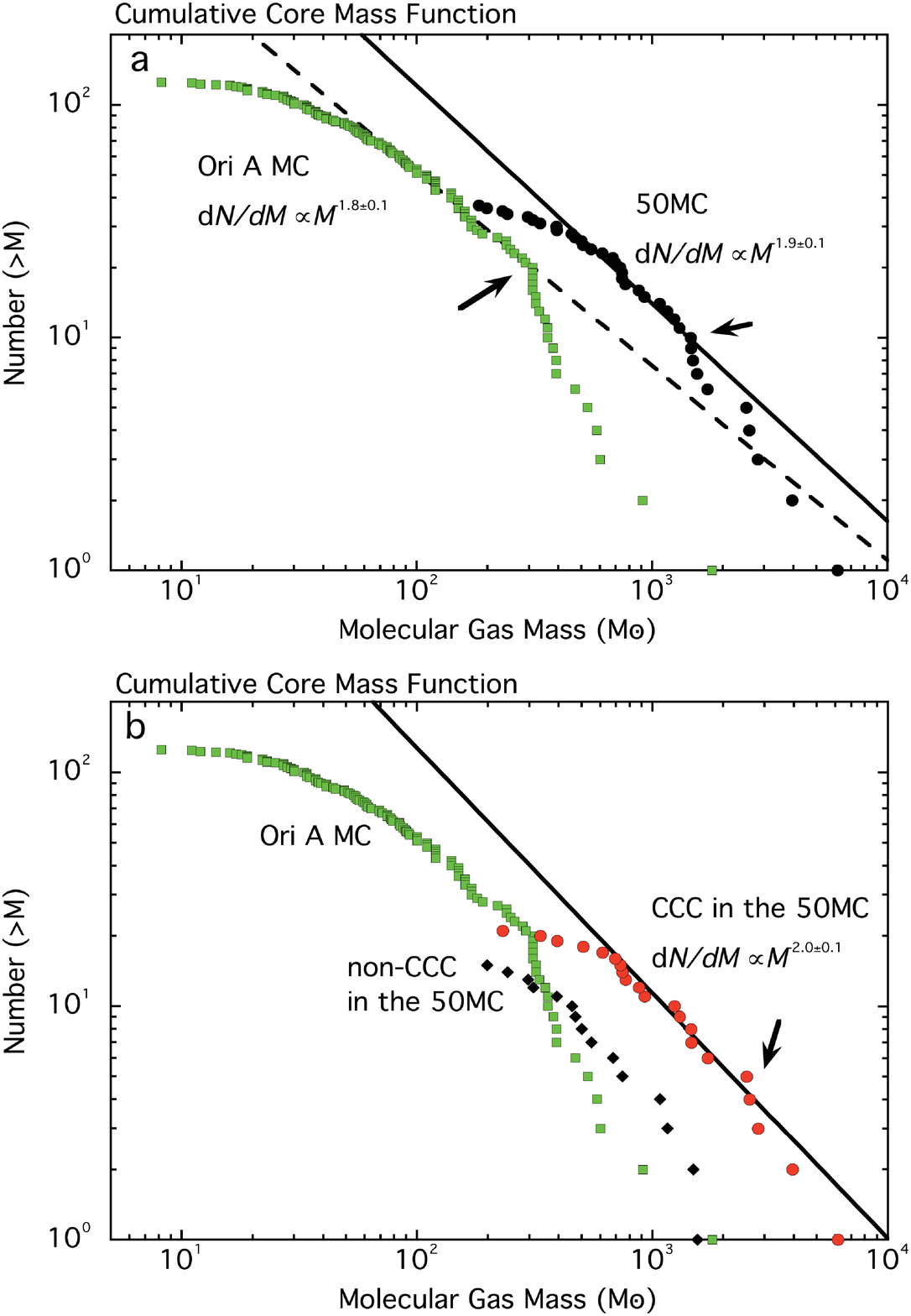}
\end{center}
\caption{{\bf a} Cumulative core mass function (CMF) of the molecular cloud cores within the 50 km s$^{-1}$ molecular cloud (50MC) is shown (black circles). The cumulative CMF of the Ori A molecular cloud is also shown  for comparison (green squares; \cite{Tatematsu}). The cumulative CMFs of the 50MC and OriAMC have breaks at $M\sim1200M_\odot$ and $M\sim300M_\odot$, respectively (black arrows). The inclined solid line shows the best-fit power law, $N\propto M^{-0.9\pm0.1}$($M>5\times 10^2$M$_\odot$). The slope corresponds to a CMF of $dN/dM\propto M^{-1.9\pm0.1}$.   
The inclined broken line shows the best-fit power law, $N\propto M^{-0.8\pm0.1}$($70M_\odot< M< 300M_\odot$). The slope corresponds to a CMF of $dN/dM\propto M^{-1.8\pm0.1}$. 

{\bf b} Cumulative CMFs of the cloud-cloud collision (CCC) (red circles) and non-CCC  (black diamonds) regions in the 50MC are shown. There is a faint break at $M\sim2500M_\odot$ in the cumulative CMFs of the CCC (black arrow). The inclined solid line shows the best-fit power law, $N\propto M^{-1.0\pm0.1}$($M>1.2\times 10^3$M$_\odot$). The slope corresponds to a CMF of $dN/dM\propto M^{-2.0\pm0.1}$.  
}
\label{Fig6}
\end{figure}
 \subsection{Core Mass Function of Molecular Cloud Cores}
Figure 6a shows  the cumulative core mass function (CMF) of the molecular cloud cores in the 50MC and also shows the cumulative CMF based on the CS $J=1-0$ emission line data of the Ori A molecular cloud (OriAMC) for comparison (\cite{Tatematsu}). 
The total mass of the 50MC is $4-10\times10^5$ M$_\odot$ derived from sub-millimeter continuum  data assuming dust temperature;$T_d=20$K and metallicity;$Z_\odot/Z=1$ (\cite{Pierce-Price2000}) . And those of CCC and non-CCC are estimated to be $3-6\times10^5$ M$_\odot$ and $1-3\times10^5$ M$_\odot$, respectively (see Fig. 3c). Meanwhile the total mass of OriAMC is $0.5-1\times10^5$ M$_\odot$ (e.g. \cite{Tatematsu}, \cite{Nagahama}, \cite{Johnston}).  Three regions have comparable total masses.

The cumulative CMF of the 50MC has a break at $M\sim1200M_\odot$ although there is other undulation (arrows in Figure 6a). We use an inflection point of the curve as the definition of ``break". The truncation in a cumulative CMF generally means that a mass of the molecular cloud cores reaches any upper limit in the mass range (e.g. \cite{Munoz}).   
Because the radius of the cores corresponding the break is larger than the beam size (see Table 1), the break is not significantly affected by the insufficient angular resolution. Meanwhile the cumulative CMF of the OriAMC has a break at $M\sim300M_\odot$. 
The inclined solid line shows the best-fit power law for the cumulative CMF of the 50MC, which is $N\propto M^{-0.9\pm0.1}$($M>500$M$_\odot$). The slope corresponds to a CMF of $dN/dM\propto M^{-1.9\pm0.1}$($M>500$M$_\odot$). The derived power law slope of the CMF is flatter than that of the 50MC, $dN/dM\propto M^{-2.6\pm0.1}$, which is previously derived with the assumption of single $T_{\mathrm{ex}}$  (\cite{Tsuboi2012b}).   
Meanwhile the slope of the best-fit power law of the OriAMC corresponds to a CMF of $dN/dM\propto M^{-1.8\pm0.1}$($70M_\odot< M< 300M_\odot$). The power law index of the 50MC is similar to that of the OriAMC. 
We applied Kolmogorov-Smirnov test to the cumulative CMFs to check whether they differ. The null hypothesis that the cumulative CMF of the 50MC is the same as that of the OriAMC is rejected because of  $p<0.001$. The CMF in the 50MC is top-heavy as compared with that in the OriAMC.

We also made cumulative CMFs of the molecular cloud cores within the CCC and non-CCC regions in the 50MC. Figure 6b shows the cumulative CMFs and the cumulative CMF  of the OriAMC for comparison.
The CMF of the CCC region is not be truncated at least up to $\sim2500M_\odot$.    On the other hand, the cumulative  CMF of the non-CCC region reaches the upper limit of $\sim1500M_\odot$.  The apparent break of the 50MC mentioned above is probably  caused by the upper limit of the non-CCC region. 
The slope of the best-fit power law of the CCC region corresponds to a CMF of $dN/dM\propto M^{-2.0\pm0.1}$($M>1200$M$_\odot$). The power law index of  the CCC region is similar to that of the OriAMC. The number of the cores including in the non-CCC region is too small to evaluate the slope of the best-fit power law accurately although the slope of the non-CCC region may be also parallel that  of the Ori AMC.
We applied Kolmogorov-Smirnov test to the CMFs to check whether the CMFs of the CCC and  non-CCC regions differ. The null hypothesis that the CMF of  the CCC region is the same as that of the non-CCC region is rejected because of $p\lesssim0.03$. 
There is a possibility that the CCC makes top-heavy CMF.

\subsection{Massive Molecular Cloud Cores in the Cloud-cloud Collision Spot}
As mentioned above, the HSF is lying between the concentration of the methanol maser at 44 GHz and the compact HII regions (see figures 3a and 3b). The age of the compact HII regions is reported to be as young as $10^4-10^5$ yr (\cite{Yusef-Zadeh2010}). Five most massive molecular cores with  $M_{\mathrm{gas}}>750M_{\odot}$ are located only around the ridge of the HSF (see figure 3d).  Less massive molecular cores are not located around the inner boundary.  The positional shift of  the methanol masers, the shocked molecular gas, the massive molecular cores, and the compact HII regions suggest that CCC makes the numerous massive molecular cores and transforms from usually observed truncated CMF to a top-heavy CMF (Cf. \cite{Inoue}, \cite{Namekata}).  The massive molecular cores produce consecutively many massive stars by accretion within $10^5$ yr (e.g. \cite{Krumholz}). 
There is a possibility that the 50 MC make a new luminous cluster as a cradle molecular cloud with CCC. In addition, such CCC-induced star formation has been reported to produce luminous clusters in the disk region (e.g. RCW49 and Westerlund 2: \cite{Furukawa}, M20: \cite{Torii}, NGC3603: \cite{Fukui}). 
On the other hand, there is no massive molecular cores with  $M_{\mathrm{gas}} >750M_{\odot}$ in the non-CCC region (see also Table 1). The region mostly corresponds to the the interacting region with the Sgr A east as mentioned previously. The interaction seems not yet to affect the core mass in the region because the age of the SNR is as young as $1\times10^4$ yr (e.g. \cite{Zhao}). 

Our observation using the Atacama Large Millimeter-submillimeter Array (ALMA) already started  to find molecular cloud cores in the cloud and image their fine structures. This information will be crucial in resolving these issues.

\section{Summary}
We performed an analysis of star-forming sites influenced by external factors, such as SNRs and/or cloud-cloud collisions,  to understand the star-forming activity in the Galactic center region using the NRO Galactic Center  Survey in SiO $v=0, J=2-1$, H$^{13}$CO$^+ J=1-0$, and CS $J=1-0$ emission lines obtained by the Nobeyama 45-m telescope.
We found a half-shell like feature with a high ratio of $R_{\mathrm{SiO/H}^{13}\mathrm{CO}^+}$ in the 50MC, which harbors an active star-forming site. 
The high ratio of $\sim6-8$ in the feature indicates that there is huge shocked molecular gas in the cloud. This feature is also seen as a half-shell feature in the position-velocity diagrams. A hypothesis explaining the chemical and kinetic properties  is that the feature is the spot of CCC.  
We analyzed the CS $J=1-0$ emission line data of the molecular cloud obtained by Nobeyama Millimeter Array to explore any influence on the star formation by CCC. The cumulative CMF in the CCC region is not truncated up to $\sim2500M_\odot$. Most massive molecular cores with  $M_{\mathrm{gas}}\gtrsim750M_{\odot}$ are located only around the ridge of the HSF.  In addition, we identified positional shift of the spot of CCC, the massive molecular cores, and the compact HII regions. If this line means a time sequence of active star formation with CCC, the 50 MC is a cradle molecular cloud forming a new luminous cluster. 

\section{Acknowledgments}
We thank Prof. Y. Kitamura at the Institute of Space and Astronautical Science and Prof. A. Habe at Hokkaido University for useful discussions. We also thank  Prof. T. Handa at Kagoshima University for useful discussions in the initial phase of this study.

%
\begin{landscape}
\begin{longtable}{lcccccccccccc}
\caption{Molecular Cloud Cores in the Galactic Center 50 km s$^{-1}$ Molecular Cloud }
\label{tbl:statistical}       
\hline
\hline
No.&$l$&$b$&$V_{\rm{LSR}}$&$\Delta V$&$R$&$M_{\rm{vir}}$&$T_{\rm{B}}$&$X(\rm{CS})$&$T_{\rm{ex}}$&$M_{\rm{gas}}$&$\alpha
$&CCC\\
&[deg.]&[deg.]&[km s$^{-1}$]&[km s$^{-1}$]&[pc]&$[M_\odot]$&[K]&&[K]&$[M_\odot]$& &[y/n]\\
\hline
\endhead
\hline

\endfoot
\hline
\endlastfoot
1&-0.0157	&-0.0700&45.0&11&0.53&1.4e+04&9.0&6.1e-08&28&9.2e+02&15&yes\\
2&-0.0117&-0.0807&29.7&9.8&0.66&1.3e+04&10&2.6e-08&55&6.1e+03&2.2&yes\\
3&-0.0104&-0.0733&45.0&11&0.55&1.5e+04&8.4&4.5e-08&32&1.5e+03&10&yes\\
4&-0.0192&-0.0696&45.0&12&0.47&1.5e+04&7.0&8.8e-08&20&3.9e+02&37&yes\\
5&-0.0290&-0.0685&37.3&7.8&0.39&4.9e+03&7.7&6.2e-08&25&3.3e+02&15&yes\\
6&-0.0149&-0.0650&37.3&14&0.38&1.5e+04&6.6&8.2e-08&21&2.3e+02&67&yes\\
7&-0.0181&-0.0565&60.3&7.1&0.17&1.8e+03&7.8&4.4e-08&26&2.0e+02&9.2&no\\
8&-0.0159&-0.0558&60.3&7.6&0.31&3.7e+03&7.7&4.5e-08&33&3.9e+02&9.5&no\\
9&-0.0232&-0.0695&41.2&7.1&0.48&5.1e+03&5.5&1.9e-08&28&1.5e+03&3.5&yes\\
10&-0.0195&-0.05741&64.1&5.6&0.16&1.1e+03&7.0&2.2e-08&28&3.1e+02&3.4&no\\
11&-0.0244&-0.0643&48.8&8.0&0.37&5.0e+03&5.8&2.1e-08&28&7.7e+02&6.5&yes\\
12&-0.0267&-0.0586&41.2&8.9&0.40&6.5e+03&7.3&3.1e-08&33&7.4e+02&8.8&no\\

13&-0.0130&-0.0606&64.1&6.2&0.29&2.4e+03&6.1&2.2e-08&30&6.1e+02&3.9&yes\\
14&-0.0067&-0.0697&45.0&9.0&0.38&6.6e+03&5.9&1.8e-08&39&1.2e+03&5.3&yes\\
15&-0.0265&-0.0545&48.8&5.7&0.41&2.8e+03&8.4&3.1e-08&23&4.7e+02&6.0&no\\

16&-0.0237&-0.0580&41.2&10&0.29&6.3e+03&6.7&3.8e-08&31&5.5e+02&11&no\\
17&-0.0161&-0.0598&56.5&4.2&0.26&9.6e+02&6.0&1.1e-08&28&7.4e+02&1.3&yes\\
18&-0.0205&-0.0600&60.3&7.6&0.34&4.1e+03&5.9&2.0e-08&28&6.9e+02&5.9&yes\\
19&-0.0224&-0.0644&52.6&7.2&0.30&3.2e+03&5.3&1.8e-08&27&5.1e+02&6.4&yes\\
20&-0.0204&-0.0560&41.2&9.3&0.39&7.0e+03&7.0&2.7e-08&35&1.2e+03&6.1&no\\
21&-0.0125&-0.0668&60.3&7.3&0.45&5.0e+03&5.1&1.4e-08&33&1.7e+03&2.9&yes\\
22&-0.0248&-0.0626&37.3&12&0.31&9.0e+03&5.4&1.0e-08&60&2.8e+03&3.2&yes\\
23&-0.0138&-0.0624&56.5&5.4&0.33&2.0e+03&5.2&9.3e-09&27&1.3e+03&1.5&yes\\
24&-0.0123&-0.0490&52.6&7.6&0.17&2.1e+03&5.6&1.3e-08&32&4.6e+02&4.5&no\\
25&-0.0182&-0.0585&64.1&4.9&0.19&9.5e+02&9.8&2.9e-08&30&3.0e+02&3.2&no\\
26&-0.0109&-0.0586&60.3&5.5&0.50&3.2e+03&5.9&1.5e-08&19&1.1e+03&3.0&no\\
27&-0.0133&-0.0516&67.9&6.7&0.32&3.1e+03&8.0&7.2e-08&26&1.8e+02&17&no\\
28&-0.0285&-0.0714&48.8&7.5&0.38&4.5e+03&4.9&1.6e-08&30&8.8e+02&5.1&yes\\
29&-0.0214&-0.0671&37.3&10&0.32&7.1e+03&4.4&8.2e-09&48&2.5e+03&2.8&yes\\
30&-0.0151&-0.0488&56.5&7.3&0.20&2.3e+03&8.9&1.1e-08&80&1.6e+03&1.5&no\\
31&-0.0022&-0.0728&75.6&4.3&0.18&6.7e+02&5.7&1.0e-08&30&6.8e+02&0.99&no\\
32&-0.0324&-0.0758& 48.8&2.4&0.19&2.4e+02&5.2&5.2e-09&26&5.0e+02&0.47&no\\
33&-0.0036&-0.0705&71.8&7.4&0.20&2.3e+03&5.1&1.6e-08&26&2.4e+02&9.6&no\\
34&-0.0083&-0.08516&14.4&9.2&0.43&7.6e+03&5.4&7.8e-09&55&3.9e+03&1.9&yes\\
35&-0.0054&-0.0813&29.7&6.9&0.25&2.5e+03&5.1&6.1e-09&55&2.6e+03&0.96&yes\\
36&-0.0053&-0.0636&56.5&8.5&0.38&5.9e+03&5.0&8.0e-09&38&1.5e+03&3.9&no\\
37&-0.0058&-0.0744&75.6&3.1&0.25&5.2e+02&4.2&5.8e-09&28&7.3e+02&0.71&yes\\
\end{longtable}
\end{landscape}


\end{document}